\documentstyle[emulateapj]{article}

\lefthead{Alonso-Herrero et al. }
\righthead{The relation between the mid-infrared emission and  
Black Hole Mass in AGN}

\begin{document}

\title{The relation between the mid-infrared emission and  
Black Hole Mass in Active Galactic Nuclei: 
A direct way to probe black hole growth?}

\begin{center}
Almudena Alonso-Herrero\footnote{Steward Observatory, 
The University of Arizona, Tucson, AZ 85721, USA}, Valentin D. Ivanov\footnote{European 
Southern Observatory, Ave. Alonso de C\'ordova No. 3107, Santiago 19001, Chile}, 
Ray Jayawardhana\footnote{Department of Astronomy, University of California at Berkeley, 
Berkeley, CA 94720, USA}, and Takashi Hosokawa\footnote{Yukawa Institute for Theoretical Physics, 
Kyoto University, Kitashirakawa, Sakyo-ku,
Kyoto 606-8502, Japan}
\end{center}

\begin{abstract}
We use a large, heterogeneous sample of local active galactic nuclei 
(AGN) that 
includes Seyfert 1s, Seyfert 2s and PG quasars to investigate for the first time the
relation between black hole mass ($M_{\rm BH}$) and mid-infrared 
nuclear emission. We find a clear relation between $M_{\rm BH}$ and 
$10\,\mu$m nuclear luminosity
for these local AGNs. There are no significant
differences between type 1 and type 2 objects, implying 
that the reprocessing of
the $10\,\mu$m nuclear emission is not severely affected by geometric and optical
depth effects. We also confirm that $M_{\rm BH}$ is related to the 
$2-10\,$keV X-ray luminosity, 
but only for the Compton thin galaxies. We present a theoretical basis 
for these empirical 
relations and discuss possible reasons for the observed scatter. Our
results show that rest-frame $10\,\mu$m and hard X-ray luminosities 
(especially the former, which is applicable to all AGN types) can 
be powerful tools
for conducting a census of BH masses at high redshift and for probing
their cosmological evolution. 

\end{abstract}


\section{Introduction}

The cosmological evolution of the black hole (BH) population is a fundamental 
property to constrain models of galaxy formation and evolution. 
In the last few years we have seen significant 
progress, especially for local galaxies where BH masses are now 
being measured.  More importantly, the discovery of two fundamental
correlations between the BH mass and the bulge luminosity 
(Kormendy \& Richstone 1995; Magorrian et al. 1998) and the BH mass and the stellar
velocity dispersion (Gebhardt et al. 2000;
Ferrarese \& Merritt 2000) suggests a tight link in the evolution 
of the BH and its galactic
host. 

To produce
reliable statistics of BH masses at high redshift, one would need 
unrealistic amounts of observing time if 
the methods employed to determine BH masses in the Local
Universe (see Section~2) were to be used. 
Thus, understanding the cosmological evolution of BHs requires
finding a good indicator of the active galactic nuclei (AGN) luminosity,  
independent from the effects of obscuration and the viewing angle, and 
directly related to the BH mass.

Among the suggested AGN power indicators are the infrared luminosity, 
the narrow line region [O\,{\sc iii}]$\lambda 5007$ line emission, the 
hard X-ray ($2-10\,$keV) emission,  
and the radio emission (e.g., Mulchaey et al. 1994). 
The $2-10\,$keV hard X-ray emission is only  
a good indicator of the intrinsic luminosity of the
AGN for those cases where it is transmitted through the torus, that is, in 
Compton thin galaxies. 

QSOs emit a significant fraction of their bolometric
luminosity in the infrared (e.g., Sanders et al. 1989).
The mid-  and far-infrared 
emission in Sys and radio quiet quasars is predominantly 
thermal in origin, namely dust emission (e.g., Rieke
1978; Barvainis 1987; McAlary \& Rieke 1988; Sanders et al. 1989; and recently
Haas et al. 2000; Polletta et al. 2000).  
Moreover, Spinoglio, \& Malkan (1989) found that the $12\,\mu$m flux is
approximately a constant fraction of the bolometric flux in 
Seyfert (Sy) galaxies and quasars.  These considerations  suggest
that  the main parameter
governing the mid-infrared behavior is the power of the central
engine, almost independently of the geometry, dust content and dust properties 
of the central region. 

For type 2 objects, the infrared emission at 
$\lambda \lesssim 5\,\mu$m may still be 
significantly affected by viewing angle effects (obscuration) and/or stellar contribution
(Alonso-Herrero et al. 2001 and references therein). At 
$\lambda \gtrsim 20\,\mu$m there may be significant contributions from star 
formation and/or the 
underlying galaxy. These contributions are not 
easily disentangled from the AGN emission unless
high spatial resolution imaging is used. 
Thus the nuclear $10\,\mu$m luminosity appears as a good  choice
to represent the AGN bolometric 
luminosity. 

In this letter we explore for the first time 
the  relation between the BH mass and  
the thermal $10\,\mu$m nuclear emission for a heterogeneous 
sample of low-redshift AGN: Sy 1s, Sy 2s, and PG quasars, 
and explore the possibility of using the mid-infrared luminosity of AGN
as a representation of their BH mass. We  
also discuss implications for the BH growth and demographics over cosmological times.

\begin{deluxetable}{lcc}
\footnotesize
\tablewidth{17cm}
\tablecaption{AGNs and BH mass methods used for 
the $10\,\mu$m luminosity vs. BH mass relation.}
\tablehead{\colhead{Method}  & \colhead{AGN} &
\colhead{Ref}}
\startdata
Reverberation Mapping & PG~0026+129$^*$, PG~0052+251$^*$, PG~0804+761$^*$, PG~0844+349$^*$, 
PG~1211+143$^*$ & 1\\
                      & PG~1226+023$^*$, PG~1229+204$^*$, PG~1307+085$^*$, PG~1351+640, 
PG~1411+442$^*$ & 1\\
                      & PG~1426+015$^*$, PG~1613+658$^*$, PG~1617+175, PG~1700+51, PG~1704+608  & 1\\
                      & PG~2130+099$^*$, Mkn~335$^*$, Mkn~590$^*$, Mkn~817, NGC~3227$^*$, 
NGC~4051$^*$  & 1\\
                      & NGC~4151$^*$, NGC~5548$^*$, NGC~7469$^*$, 3C390.3$^*$, Akn~120$^*$, 
F~9$^*$, IC~4329A$^*$, 
Mkn~509$^*$   & 1    \\
                      & Mkn~279$^*$, NGC~3516$^*$, Mkn~841$^*$, Mkn~766$^*$, NGC~4593$^*$ & 2\\
$M_{\rm BH}$ -- $\sigma_*$ relation &  
                      NGC~1386, NGC~1566, NGC~1667, 
                     NGC~2110$^*$, NGC~3185, NGC~3362, NGC~3982 & 3, 4 \\ 
                     &NGC~4388$^*$, NGC~4579$^*$, NGC~5252$^*$, NGC~5273, NGC~5283, NGC~5347 & 3, 4 \\
                     & NGC~5695, NGC~5929, NGC~5940, NGC~5953, NGC~6104 , NGC~7682 & 3, 4\\  
                     & NGC~7674, Mkn~1, Mkn~3$^*$, Mkn~348$^*$, Mkn~530, Mkn~573, Mkn~744, Mkn~1040 & 
3, 4\\   
Stellar/Gas kinematics & NGC~1386, NGC~4258$^*$, NGC~1068, Circinus, NGC~4945$^*$ 
& 5\\
                     & NGC~4395$^*$ (upper limit) & 6
\enddata
\tablecomments{The galaxies marked with $^*$ were used for the hard X-ray luminosity
vs. BH mass relation (Figure~1, right panel).\\
References for $M_{\rm BH}$ and method: 1. Kaspi et al. (2000). 2. Compilation of Wandel (2002).
3. Nelson \& Whittle (1995). 4. Ferrarese \& Merritt (2000). 5. Compilation of
Moran et al. (1999). 6. Filippenko \& Ho (2001).\\
References for the $10\,\mu$m fluxes. PG quasars: 
Neugebauer et al. (1987) and Sanders et al. (1989). 
Sy galaxies: small aperture ground-based measurements from Rieke (1978); Maiolino et al. (1995);
Krabbe, B\"oker, \& Maiolino (2001); P. Lira (2001, private communication), and {\it ISO} fluxes 
from Clavel et al. (2000). }
\end{deluxetable}

\section{The Data and Correlations}

We have compiled BH masses of local AGNs -- Sy galaxies and 
PG quasars --  derived using three methods and 
for which nuclear $10\,\mu$m fluxes are available (see Table~1 
for the sample and references for the data). The most 
direct way is from spatially resolved stellar kinematics (see 
Ho 1999 for a review), and 
measurements of masers in disks around massive 
BHs (see Moran, Greenhill, \&  
Herrnstein 1999 for a review). 

In the case of bright active galaxies
where the stellar emission is swamped by the AGN 
emission, the best technique to derive BH masses is reverberation mapping  
(see a review by Peterson 1993).  For sample of AGNs analyzed in  
Kaspi et al. (2000) we have used the BH masses (and errors) derived from the mean of 
the FWHM velocity measurements.

The third and most indirect method uses the  empirical relation between the 
BH mass and the stellar velocity dispersion (the $M_{\rm BH}$ -- 
$\sigma_*$ relation)
found by Ferrarese \& Merritt (2000): $\log M_{\rm BH} = 4.80(\pm0.54)\,
\log\sigma_*-2.9(\pm1.3)$. The $M_{\rm BH}$ errors here are dominated by the empirical
correlation uncertainty rather than by the observational errors
in $\sigma_*$.  In Figure~1 we only plot the errors associated with 
the slope of the relation, that is, $\Delta \log M_{\rm BH} = 0.54\times
\log\sigma_*$. 

We also explore the $2-10$\,keV hard X-ray luminosity vs. BH mass relation 
for Compton thin 
AGNs based upon the good correlation found between the 
hard X-ray and 
the $10\,\mu$m luminosities in Sys and PG quasars 
(Alonso-Herrero et al. 2001, Krabbe et al. 2001). 
The $2-10\,$keV hard X-ray luminosities (corrected for absorption) are from 
Nandra \& Pounds (1994), Lawson \& Turner (1997), Bassani et al. (1999), 
George et al. (2000), Iwasawa et al. (2000), 
and the compilation in Alonso-Herrero et al. (1997). 
We have included a few low luminosity AGNs with available hard X-ray fluxes corrected for 
absorption (although   
mid-infrared fluxes were not available). These are NGC~3031, 
NGC~4594, NGC~6251 ($M_{\rm BH}$ 
from the compilation of Ho 2002), NGC~3079 ($M_{\rm BH}$
from Moran et al. 1999) and NGC~5194 ($M_{\rm BH}$ derived from 
the $M_{\rm BH}$ -- $\sigma_*$ relation, where $\sigma_*$ is from Nelson \&
Whittle 1995). 

When available we have used distances derived from surface 
brightness fluctuations (Tonry et al. 2001). 
For the rest of the galaxies distances were computed 
using $H_0= 75\,{\rm km\,s}^{-1}\,{\rm Mpc}^{-1}$.

\begin{figure*}
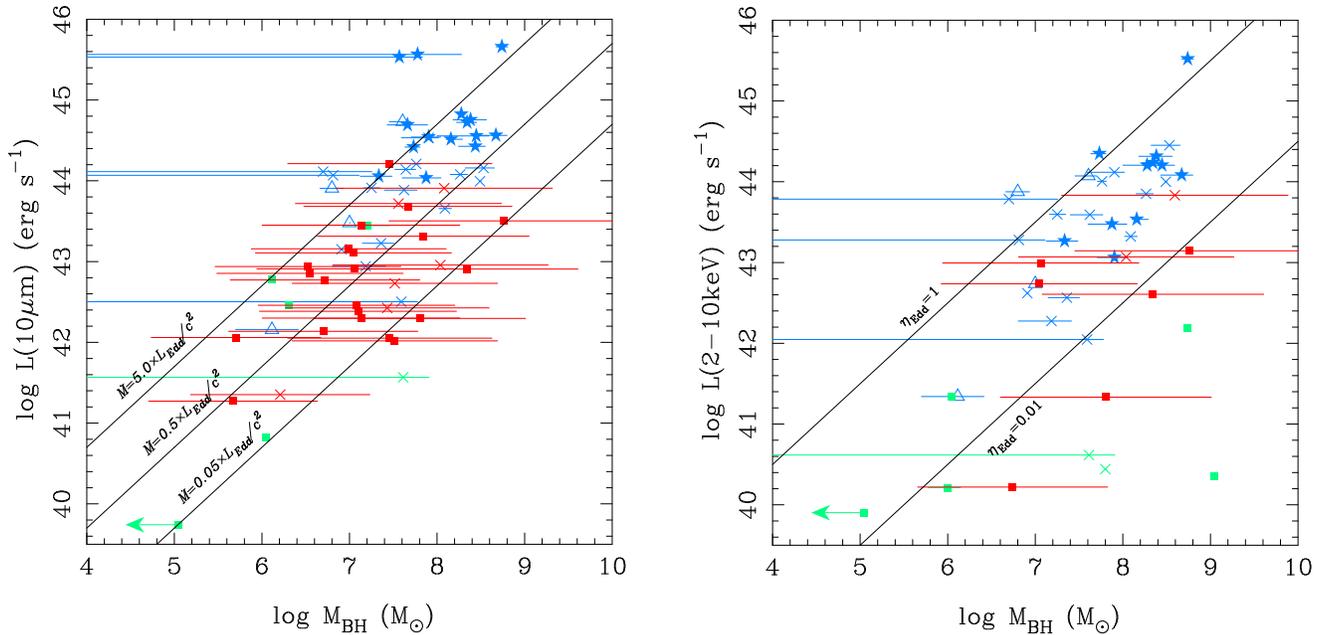

\figurenum{1}
\plotfiddle{figure1a_color.ps}{425pt}{-90}{55}{55}{-270}{505}
\plotfiddle{figure1b_color.ps}{425pt}{-90}{55}{55}{-10}{940}
\vspace{-22cm}
\caption{{\it Left panel:} Relation between the 
$10\,\mu$m nuclear luminosity and the BH mass. Filled stars are PG quasars, X's are 
Sy 1s, open triangles are NLS1s, and filled squares are Sy 2s.
In the on-line version of this figure the data are color coded 
according to the method used for determining the BH mass: green from stellar/gas 
kinematics, 
blue from reverberation mapping and red from the $M_{\rm BH}$ -- $\sigma_*$ 
relation. {\it Right panel:} same as left 
panel but for the hard X-ray ($2-10\,$keV) luminosity and BH mass.}
\end{figure*}

\section{Discussion and Conclusions}

The relations between the BH mass and 
the $10\,\mu$m nuclear luminosity and $2-10\,$keV luminosity for our AGN sample 
are presented in Figure~1 (left and right panel respectively). 
The latter relation has been previously analyzed by Wandel \& Mushotzky (1986) and 
Awaki et al. (2001) for a limited number of AGNs.
Ho (2002) and Wu \& Han (2001) have investigated the relationship between the  
radio nuclear and total power and the BH mass and found a good correlation for 
bright AGNs. Ho (2002) however finds that 
weakly active galaxies do not follow the same relation.

In a series of papers, McLeod \& Rieke (1995), McLeod, Rieke, \&
Storrie-Lombardi (1999) and McLeod \& McLeod (2001) have demonstrated that there is
a relation between the nuclear $B$ magnitude of quasars and the near-infrared luminosity of 
the host galaxy (that is, the stellar mass of the galaxy). 
This relation is also present in luminous infrared galaxies when the far-infrared
luminosity is used to represent the bolometric luminosity (McLeod et al. 1999).
McLeod \& McLeod (2001) interpreted these relations as the maximum nuclear 
luminosity possible 
in a host galaxy of a given near-infrared luminosity (mass). 
In view of this relation and the good correlation 
between the BH mass and the bulge luminosity of the host galaxy it 
would not be unreasonable to expect   
a relation between the BH mass and the AGN luminosity represented 
by the $10\,\mu$m or the hard X-ray luminosities, the latter only for 
Compton thin objects.

In the hard X-ray vs. $M_{\rm BH}$ relation the
 lines ($\eta_{\rm Edd} =0.01$ and $\eta_{\rm Edd} =1$) are the Eddington
ratios computed in  Awaki et al. (2001, their figure~3) 
assuming a bolometric correction of $\frac {L_{\rm bol}}{L_{\rm X}}=27.2$ (note 
that these authors pointed out that such bolometric correction may be lower
for low luminosity AGNs). All the objects located to the right of the 
$\eta_{\rm Edd} =0.01$ line are low-luminosity AGNs. 
Although this is a good relation, it is only useful for Compton 
thin galaxies (column densities $N_{\rm H} \le 10^{24}\,{\rm cm}^{-2}$). 
Unfortunately for a significant fraction of type 2 objects the $2-10\,$keV 
X-ray emission is not transmitted through the torus (e.g., Bassani et al. 
1999), and the $2-10\,$keV luminosities cannot be used to trace 
the obscured BH growth.

The scatter in the relations may be accounted for by 
efficiency variations in the accretion process in different kinds of AGN
and/or evolutionary effects. The evolution will be relevant if the 
accretion rate is higher in the early growing stage of an AGN, as proposed by 
Mathur, Kuraszkiewicz, \& Czerny (2001) to explain the properties 
of narrow line Sy 1 galaxies (NLS1s). This class of galaxies would  
represent an early stage of the AGN evolution based upon their 
lower  BH to bulge mass ratios. The  
four NLS1s (shown as open triangles in Figure~1) 
in our sample show a similar behavior to the other Sy galaxies in 
both panels of Figure~1, if only with slightly higher accretion rates than the average. 
However the small number of NLSy1s does not allow us to reach a firm conclusion.

 Another factor that may introduce some scatter in the mid-infrared 
vs. BH mass relation could be the presence of 
nuclear star formation in Sy 2 galaxies. If star formation were present 
(especially for those galaxies observed with the relatively large 
{\it ISO} aperture), 
we would expect an excess of $10\,\mu$m emission for a given BH mass. We do not
find such excesses in our sample of galaxies. On the other hand, if the 
optical depths were high for type 2 objects even at $10\,\mu$m, 
then for a given BH mass they should lie below type 1 objects. 
Type 1 objects do not generally lie above
type 2 objects in the $10\,\mu$m luminosity vs. BH mass diagram, 
suggesting that the reprocessing of
the $10\,\mu$m nuclear emission is not severely affected by geometric and optical
depth effects (see also Alonso-Herrero et al. 2001).

Hosokawa et al. (2001) and Kawaguchi, Shimura, \&
Mineshige (2001) constructed a disk-corona model to account for the 
X-ray through infrared spectral energy distributions (SEDs) of quasars and Sy galaxies.
These models demonstrate that the AGN SEDs are sensitive to both  
the BH mass ($M_{\rm BH}$) 
and accretion rate ($\dot M$). In the optical, UV and soft X-ray spectral ranges  
decreasing $\dot M$ and/or increasing $M_{\rm BH}$ will move the peak 
of the SED toward longer wavelengths 
(i.e., figure~1 in Hosokawa et al. 2001).
On the other hand, the hard X-ray and mid-infrared  spectral 
shapes remain approximately constant for varying $M_{\rm BH}$ and 
$\dot M$. This is expected, as in this model the infrared bump is produced 
by dust heated by the AGN, and modelled with constant spectral indices, 
regardless of  $M_{\rm BH}$ or $\dot M$. The infrared power varies so that  
the luminosity ratio of the big blue bump to the infrared bump remains 
constant.

The predicted $10\,\mu$m luminosity as a  
function of  $M_{\rm BH}$ from Hosokawa et al. 
(2001) models is shown in Figure~1 (left panel) 
for two accretion rates\footnote{For reference in this model a 
BH with  an accretion rate of $\dot M =12 \times 
L_{\rm Edd}/c^2$ will represent a disk with Eddington luminosity, 
$L_{\rm Edd} = 1.5 \times 
10^{38}\frac{M_{\rm BH}}{{\rm M}_\odot}$ (in erg s$^{-1})$
$L_{\rm Edd}$.}  $\dot M =5-0.5 \times 
L_{\rm Edd}/c^2$. The third accretion rate $\dot M =0.05 \times 
L_{\rm Edd}/c^2$ curve has been extrapolated from the 
above mass accretion rates.
The $10\,\mu$m luminosities of PG quasars are well reproduced
with accretion rates of  $\dot M =0.5-5 \times 
L_{\rm Edd}/c^2$ (as found also by Kawaguchi
et al. 2001 from the optical to X-ray properties), whereas Sy galaxies in
general require lower accretion rates (see also e.g., Su \& Malkan 1989).

\begin{figure*}
\figurenum{2}
\plotfiddle{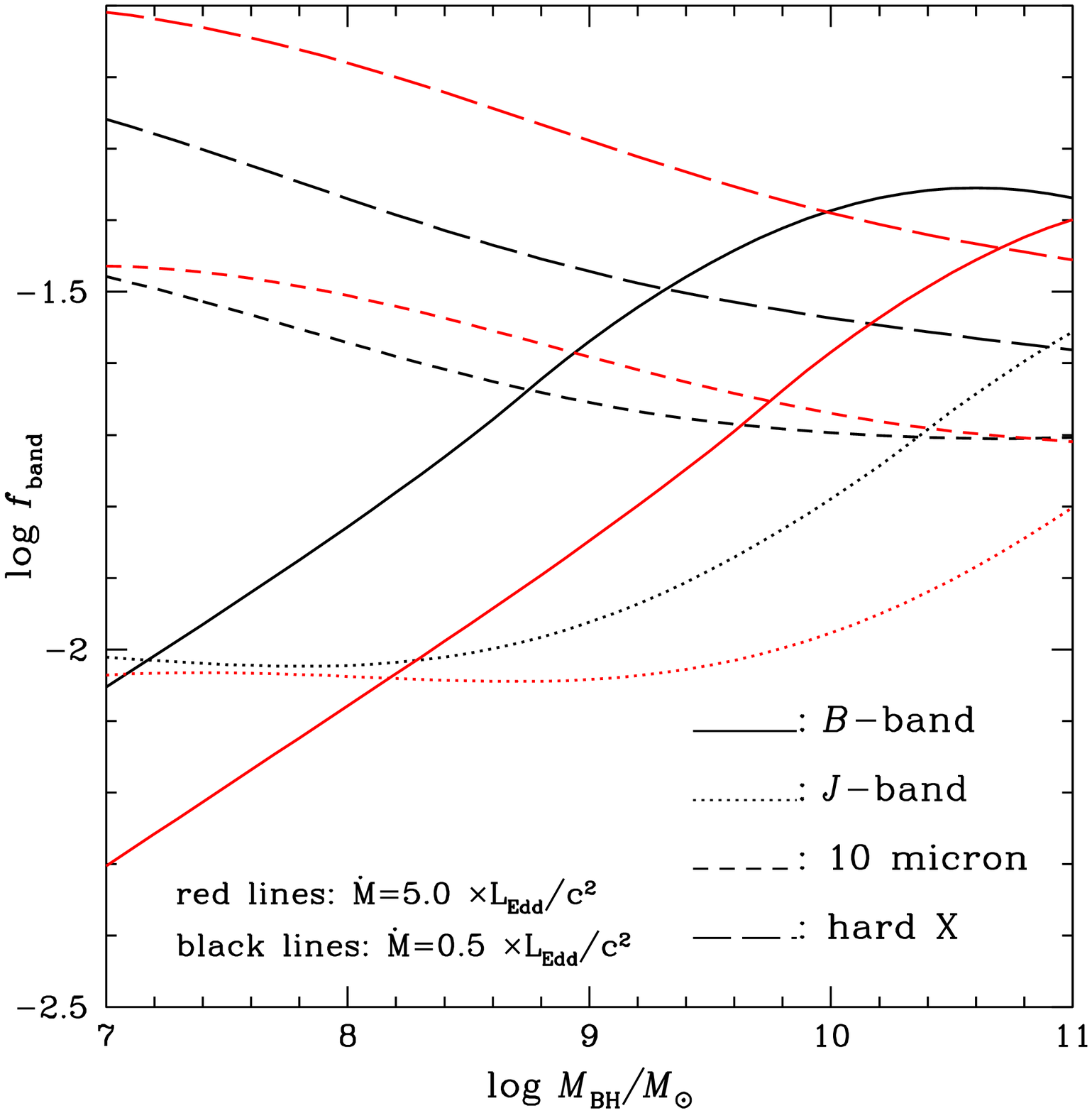}{425pt}{0}{35}{35}{-230}{170}
\plotfiddle{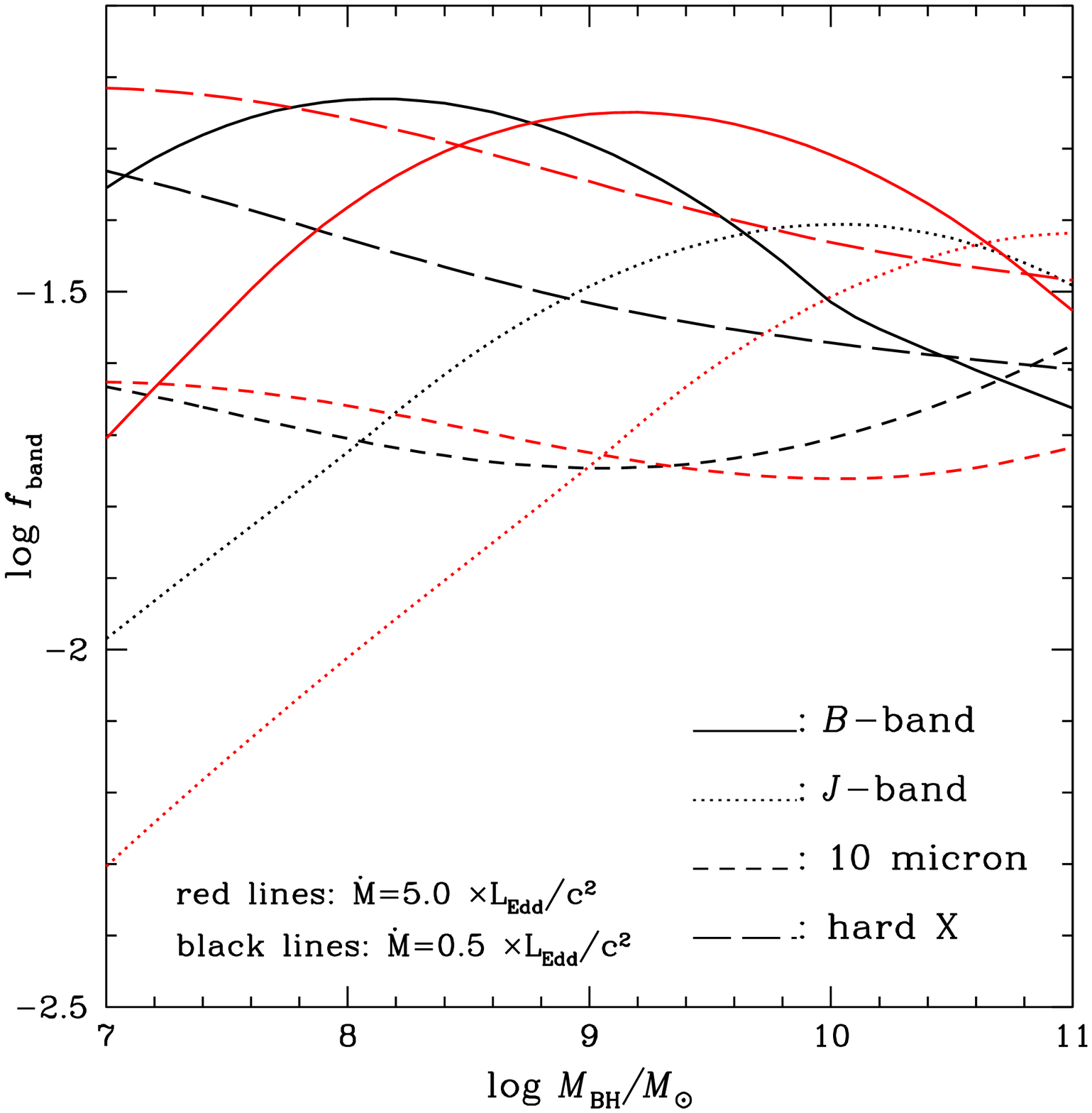}{425pt}{0}{35}{35}{-0}{605}
\vspace{-23cm}
\caption{Predictions of Hosokawa et al. (2001) disk corona model for 
the ratio of the $B$-band, $J$-band ($1.2\,\mu$m), $N$-band ($10\,\mu$m) and 
hard X-ray ($2-10\,$keV) luminosities to 
the bolometric luminosity of the AGN as a function of $M_{\rm BH}$ at $z=0$ (left panel) 
and $z=3$ in the observer's rest-frame (right panel). The 
accretion rate is  $\dot M = 5\times 
L_{\rm Edd}/c^2$ (thin lines, red lines in the color version) and $\dot M = 0.5\times 
L_{\rm Edd}/c^2$ (thick lines, black lines in the color version).}
\end{figure*}

Figure~2 shows the ratio of the optical $B$-band, 
near-infrared $J$-band, $10\,\mu$m and hard X-ray to the AGN bolometric luminosity 
as a function of $M_{\rm BH}$ at two different redshifts $z=0$ (left panel) and $z=3$ 
(in the observer's rest-frame, right panel) 
from Hosokawa et al. (2001) model predictions.  The accretion  
rate are  $\dot M =5.0 \times L_{\rm Edd}/c^2$  and 
$\dot M =0.5 \times L_{\rm Edd}/c^2$. 
From this figure it is clear that the ratios of the $10\,\mu$m and hard X-ray luminosities to 
the AGN bolometric luminosity remain  approximately constant for increasing $M_{\rm BH}$
at a given $\dot M$ over redshifts of $z=0-3$. Moreover, the 
ratio of the $10\,\mu$m luminosity to the AGN bolometric luminosity is also almost constant 
for the two accretion rates. This shows that 
the $10\,\mu$m luminosites can be used to trace the BH 
growth by comparing the mid-infrared luminosity functions at different epochs. 

The optical to bolometric luminosity ratio varies 
significantly as a function of the BH mass for redshifts $z=0$ and $z=3$, 
and the two accretion rates. Although the $J$-band 
to bolometric luminosity ratio does only vary slightly with 
$M_{\rm BH}$ at $z=0$ and $\dot M =5.0 \times L_{\rm Edd}/c^2$, it 
does significantly at $z=3$ for both accretion rates.
Leaving aside the unavoidable obscuration  (viewing angle) effects of type 2 AGNs,  
the complex dependence of the UV, optical and near-infrared SEDs with $M_{\rm BH}$ 
and $\dot M$ would make it more intricate the interpretation of the 
time evolution of luminosity functions
to infer the BH growth.

There is now growing evidence for a missing population of obscured type 2 AGNs,
necessary to reproduce 
the hard X-ray background (e.g., Fabian \& Iwasawa 1999), as well as to 
reconcile the observed local comoving density of BHs with that 
predicted from the QSO luminosity function at $z = 3$ (Haehnelt, 
Natarajan, \& Rees 1998).  This indicates that 
a significant fraction of the accretion by BHs may be 
obscured by dust. The two relations discussed here suggest that the
hard X-ray luminosity  and more importantly, the mid-infrared luminosity can be effective 
tools to probe the processes taking place in obscured nuclei of galaxies.
These relations can be used to conduct a detailed
census of the BH masses over a range of redshifts, and measure
directly their growth rate as long as an estimate of the accretion rate is available.
The present X-ray space missions are already producing 
results (e.g., Brandt et al. 2002 and references therein).
The launch of {\it SIRTF} will provide further capabilities to obtain
$M_{\rm BH}$ for obscured AGNs at redshifts of up to $z\simeq 4$.

Summarizing, for a heterogeneous sample of local AGNs we find: (i) the $M_{\rm BH}$ correlates 
with the $10\,\mu$m nuclear luminosity in local Sy galaxies and PG quasars; 
(ii) the available data suggest no significant differences
between type 1 and type 2 objects, implying that the reprocessing of
the $10\,\mu$m nuclear emission is not severely affected by geometric and optical
depth effects;
(iii) there is a good relation between the $M_{\rm BH}$ and the 
$2-10\,$keV hard X-ray luminosity, but only for Compton thin galaxies; 
(iv) the $M_{\rm BH}$ vs. $L(10\,\mu{\rm m})$ relation will allow to 
estimate the obscured BH mass growth over cosmological 
times using future observations with {\it SIRTF}.

We thank Dave Alexander, Roberto Maiolino and Laura Ferrarese 
for helpful discussions, and Paulina 
Lira for providing us with data prior to publication.
We are also grateful to an anonymous referee for useful comments.
AAH acknowledges support from NASA 
Contract 960785 through the Jet Propulsion Laboratory.

\end{document}